\documentclass[a4paper,11pt]{article}
\usepackage{graphicx}
\usepackage{cite}
\usepackage{amsfonts}
\usepackage{amssymb}
\usepackage{latexsym}
\usepackage{url}
\usepackage{color}
\usepackage{ntheorem}

\usepackage{float} % I think this allows fixation of graphics to certain locations using [H!]-like commands
\usepackage{xcolor} % see color package
\usepackage{graphicx} % takes care of graphic including machinery
\usepackage[normalem]{ulem}
\usepackage{bbm}
\usepackage{dsfont}
\usepackage[mathscr]{euscript}
\usepackage{ntheorem}

\usepackage{pstricks,pst-node,pst-coil,pst-plot}

\newcommand{\mnotex}[1]%{}
{\protect{\stepcounter{mnotecount}}$^{\mbox{\footnotesize
$%\!\!\!\!\!\!\,
\bullet$\themnotecount}}$ \marginpar{%\color{red}%
\raggedright\tiny\em
$\!\!\!\!\!\!\,\bullet$\themnotecount: #1} }

\newcommand{\jamesx}[1]{}%{\james{ {#1}}}
\renewcommand{\jamesx}[1]{{\mnote{{\color{blue}{\bf jg:}
#1} }}}

\newcommand{\cref}[1]{\mbox{{\color{red}FIXME; what is the 4 for?}}4\emph{\ref{#1})}}

\newcommand{\h}[2]{#1\dotfill\ #2\\\ptc{fixme}}
\newcommand{\eqref}[1]{\eq{#1}}

\newcommand{\myX}{{\blue{X}}}
\newtheorem{theorem}{Theorem}[section]

\newtheorem{Theorem}[theorem]{\sc  Theorem\rm}

\newtheorem{Lemma}[theorem]{\sc Lemma\rm}

\newtheorem{Proposition}[theorem]{\sc Proposition\rm}

\newcommand{\hs}{\cH_{\mbox{\scriptsize sing}}}

\newcommand{\beadl}[1]{\begin{deqarr}\label{#1}}
\newcommand{\eeadl}[1]{\arrlabel{#1}\end{deqarr}}%
\def\nz{\ifmmode {I\hskip -3pt N} \else {\hbox {$I\hskip -3pt N$}}\fi}
\def\zz{\ifmmode {Z\hskip -4.8pt Z} \else
       {\hbox {$Z\hskip -4.8pt Z$}}\fi}
\def\qz{\ifmmode {Q\hskip -5.0pt\vrule height6.0pt depth 0pt
       \hskip 6pt} \else {\hbox
       {$Q\hskip -5.0pt\vrule height6.0pt depth 0pt\hskip 6pt$}}\fi}
\def\rz{\ifmmode {I\hskip -3pt R} \else {\hbox {$I\hskip -3pt R$}}\fi}
\def\cz{\ifmmode {C\hskip -4.8pt\vrule height5.8pt\hskip 6.3pt} \else
       {\hbox {$C\hskip -4.8pt\vrule height5.8pt\hskip 6.3pt$}}\fi}
\def\au{{\setbox0=\hbox{\lower1.36775ex\hbox{''}\kern-.05em}\dp0=.36775ex\hs
kip0pt\box0}}
\def\ao{{}\kern-.10em\hbox{``}}

\newcommand\Gregbeq{\begin{eqnarray}}
\newcommand\Gregeeq{\end{eqnarray}}

\def\cH{{\cal H}}

\def\h1{{\hat 1}}
\def\h2{{\hat 2}}

\def\3f{\frac{3}{2}}

\newcommand{\roscoff}[1]{}

{\catcode `\@=11 \global\let\AddToReset=\@addtoreset}
\AddToReset{figure}{section}

\DeclareFontFamily{OT1}{rsfs}{}
\DeclareFontShape{OT1}{rsfs}{m}{n}{ <-7> rsfs5 <7-10> rsfs7 <10-> rsfs10}{}
\DeclareMathAlphabet{\mycal}{OT1}{rsfs}{m}{n}

{\catcode `\@=11 \global\let\AddToReset=\@addtoreset}
\AddToReset{equation}{section}

\newcounter{mnotecount}[section]

\renewcommand{\themnotecount}{\thesection.\arabic{mnotecount}}

\newcommand{\jlcax}[1]{}
\newcommand{\eean}{\nonumber\end{eqnarray}}

\newcommand{\kk}[1]{}%{\mnote{{\bf If we consider the KK case:} #1}}

\newcommand{\mcH}{{\mycal H}}

\newcommand{\beq}{\begin{equation}}

\newcommand{\FS}       %{F_1} %
                  {F}
\newcommand{\HS} %{F_2}
       {H_{\mbox{\scriptsize volume}}}

{\ptc{this should be removed in the oberwolfach version}}%

\newcommand{\eeal}[1]{\label{#1}\end{eqnarray}}
\newcommand{\bed}{\begin{deqarr}}
\newcommand{\eed}{\end{deqarr}}
\newcommand{\bedl}[1]{\begin{deqarr}\label{#1}}
\newcommand{\eedl}[2]{\arrlabel{#1}\label{#2}\end{deqarr}}

\newcommand{\bel}[1]{\begin{equation}\label{#1}}
\newcommand{\bea}{\begin{eqnarray}}
\newcommand{\bean}{\begin{eqnarray}\nonumber}
\newcommand{\beal}[1]{\begin{eqnarray}\label{#1}}
\newcommand{\eea}{\end{eqnarray}}

\newcommand{\qedskip}{\hfill $\Box$\medskip}
\newcommand{\proof}{\noindent {\sc Proof:\ }}
\newcommand{\be}{\begin{equation}}
\newcommand{\eeq}{\end{equation}}
\newcommand{\ee}{\end{equation}}
\newcommand{\beqa}{\begin{eqnarray}}
\newcommand{\eeqa}{\end{eqnarray}}
\newcommand{\beqan}{\begin{eqnarray*}}
\newcommand{\eeqan}{\end{eqnarray*}}
\newcommand{\ba}{\begin{array}}
\newcommand{\ea}{\end{array}}

\newcommand{\hyp}{\mycal S}

\newcommand{\mcM}{{\mycal M}}

\newcommand{\mnote}[1]%{}
{\protect{\stepcounter{mnotecount}}$^{\mbox{\footnotesize
$%\!\!\!\!\!\!\,
\bullet$\themnotecount}}$ \marginpar{%\color{red}%
\raggedright\tiny\em
$\!\!\!\!\!\!\,\bullet$\themnotecount: #1} }

\newcommand{\warn}[1]%{}%{}
{\protect{\stepcounter{mnotecount}}$^{\mbox{\footnotesize
$%\!\!\!\!\!\!\,
\bullet$\themnotecount}}$ \marginpar{%\color{red}%
\raggedright\tiny\em
$\!\!\!\!\!\!\,\bullet$\themnotecount: {\bf Warning:} #1} }

\newcommand{\N}{\mathbb N}

\newcommand{\eq}[1]{(\ref{#1})}

\newcommand{\ptc}[1]{\mnote{{\bf ptc:}#1}}

\newcommand{\beqar}{\begin{deqarr}}
\newcommand{\eeqar}{\end{deqarr}}

\newcommand{\beaa}{\begin{eqnarray*}}
\newcommand{\eeaa}{\end{eqnarray*}}

\newcommand{\doc}{\langle\langle \mcM\rangle\rangle}
\newcommand{\bethm}{\begin{Theorem}}
\newcommand{\et}{\end{Theorem}}
\newcommand{\bl}{\begin{Lemma}}

\DeclareFontFamily{OT1}{rsfs}{}
\DeclareFontShape{OT1}{rsfs}{m}{n}{ <-7> rsfs5 <7-10> rsfs7 <10-> rsfs10}{}
\DeclareMathAlphabet{\mycal}{OT1}{rsfs}{m}{n}

{\catcode `\@=11 \global\let\AddToReset=\@addtoreset}
\AddToReset{equation}{section}

{\catcode `\@=11 \global\let\AddToReset=\@addtoreset}
\AddToReset{figure}{section}

{\catcode `\@=11 \global\let\AddToReset=\@addtoreset}
\AddToReset{table}{section}

\renewcommand{\themnotecount}{\thesection.\arabic{mnotecount}}

\renewcommand{\blue}[1]{#1}
\renewcommand{\red}[1]{#1}

\begin{document}

\title{Remarks on stationary vacuum black holes%
\thanks{Preprint UWThPh-2023-18}}
\author{
Piotr T. Chru\'sciel\thanks{Gravitational Physics, University of Vienna, Boltzmanngasse 5, 1090 Vienna, Austria} {}\thanks{Email  {Piotr.Chrusciel@univie.ac.at}, URL {
http://homepage.univie.ac.at/piotr.chrusciel}}
%\\ MK \thanks{Email  xx}  \vspace{0.5em}\\
 }
\maketitle

%\vspace{-0.2em}

\begin{abstract}
We finish the proof of the no-hair theorem for  stationary, analytic, connected, suitably regular, four dimensional vacuum black holes.
We show how to define the surface gravity and the angular velocity of horizons without assuming analyticity. We point out that, under the usual regularity conditions, vacuum near-horizon geometries are Kerrian without assuming analyticity.
\end{abstract}

\noindent
\hspace{2.1em} PACS numbers: 04.20.Dw, 02.40.-k

\tableofcontents

\section{Introduction}

Recall that a  black hole is called \emph{degenerate} if there exists a Killing vector field tangent to the generators of the horizon which has vanishing surface gravity. A stationary black hole is called \emph{rotating} if the stationary Killing vector is \emph{not} tangent to the generators of the horizon.

The theory of stationary black holes  developed by several authors, summarised in \cite{ChCo}, shows that the exteriors of  sufficiently regular, analytic, connected vacuum
black holes which are \emph{either rotating, or non-rotating and non-degenerate}, are modeled by the Kerr metric.
 In this note we point out that, under the remaining conditions, the hypothesis of \emph{non-degeneracy} is not needed (see~\cite{ChCo} for definitions), so that it holds:

\begin{Theorem}
\label{T12X22.1a}
The domain of outer communications of a stationary,  vacuum, analytic, four dimensional, asymptotically flat and $I^+$-regular, connected black hole is isometrically diffeomorphic to one in a Kerr spacetime.
\end{Theorem}

In a nutshell, Theorem~\ref{T12X22.1a} follows from~\cite[Theorem~5.1]{ChAscona}  together with the results in~\cite{ChNguyen,ChCo}  and the observation below  that   \emph{at least one component of the horizon must be rotating if all are mean-degenerate}.

We note that this last observation is an obvious consequence of Smarr-type mass formulae of \cite{SudarskyWald93}. However, in \cite{SudarskyWald93} it is assumed   at the outset that the isometry group of the spacetime is at least two dimensional; here this is proved rather than assumed when the metric is analytic.
 While this is not needed for the uniqueness result above, we  show that these mass formulae can be established without assuming existence of two Killing vectors in any case.

Last but not least, we point out that the near-horizon geometry of four-dimensional degenerate vacuum black holes in $I^+$-regular spacetimes arises from an extreme Kerr metric without assuming analyticity.

\section{Analytic spacetimes}

Let $\myX $ denote the Killing vector which generates time translations in the asymptotically flat region.
By definition of $I^+$-regularity,
 the domain of outer communications $\doc$ contains an asymptotically flat hypersurface $\hyp$ with compact boundary $\partial \hyp$ lying on a non-empty future event horizon.  The existence of   a three-dimensional hypersurface $\hyp$ with this topology implies in particular that the ADM mass of $\hyp$ is positive~\cite{Herzlich:mass}.
While Theorem~\ref{T12X22.1a} assumes connectedness, in what follows we will allow the future event horizon to have more than one component, with the connected components of $\partial \hyp$  denoted by $\partial \hyp_{\red{(i)}}$, where $i\in \{1,\ldots,N\}$ for some $N\in \N$.

Suppose, first, that all components  of the horizon are non-rotating. By definition,  the stationary Killing vector is then tangent to the generators of each horizon, which allows one to define the surface gravity there.

Since the ADM mass $m$
of $\hyp$  coincides with the Komar integral~\cite{BeigKomar} we have
%\ptc{(compare~\cite[Equation~(4.6.24)]{ChBlackHoles})}
% \ptcr{coeffs? sign }
%
\begin{eqnarray}
   {8\pi} m
    & = &   -\int_{S_\infty} \nabla ^\mu X^\nu dS_{\mu\nu}
    =
    -\int _{\partial \hyp } \nabla ^\mu X^\nu dS_{\mu\nu}
    =
    2 \sum_{\red{(i)}} \int _{\partial \hyp_{\red{(i)}} } \kappa_{\red{(i)}} \, d \mu
     \nonumber
 \\
    &   = &
    2 \sum_{\red{(i)}} \kappa_{\red{(i)}} A_{\red{(i)}}
    \,,
 \label{20IX22.1}
\end{eqnarray}
where $d\mu$ is the  measure induced by $g$  on $\partial_\hyp$ and  $\kappa_{\red{(i)}}\ge0 $ is the (constant) surface gravity of $\partial \hyp_{\red{(i)}}$
 and $A_{\red{(i)}}$ its area. If all the components are degenerate, i.e.\ all the $\kappa_{\red{(i)}}$'s are zero, one obtains a contradiction with the positive energy theorem.

We conclude that at least one $\kappa_{\red{(i)}}$ has to be non-zero and, if not, at least one component must be rotating.

If \emph{all}  $\kappa_{\red{(i)}}$'s are non-zero
we have the \emph{staticity theorem}: there is only one component, the d.o.c.\ is static, and the black hole is Schwarzschild~\cite{Sudarsky:wald,ChWald} (compare~\cite{ChCo}).
We note that in this situation the hypotheses of analyticity and  connectedness are not needed.

From what has been said we see that, under the conditions of Theorem~\ref{T12X22.1a}, where a connected black-hole has been assumed at the outset, either the black hole is Schwarzschild or the horizon is rotating. Since we assumed analyticity, the proof of  Theorem~\ref{T12X22.1a} is completed by the main theorem of~\cite{ChNguyen}.

We note that  properties alternative to full analyticity would suffice: one-sided analyticity at the horizon (which can be established in some situations~\cite{ChAPP,BCM}), or unique continuation (cf.~\cite{AIK} and references therein).

We also note that  the argument does not work for charged black holes, in which case \eqref{20IX22.1} is modified by a contribution from the Maxwell field.

\section{Without analyticity}
 \label{s30IV23.1}

Given an \emph{axisymmetric and stationary}, possibly multi-component black-hole spacetime, the standard way to define the surface gravity and the angular velocity of each component of the horizon proceeds as follows: Let $X$ denote the Killing vector field which defines stationarity (for instance, for asymptotically flat configurations, this is the Killing vector field which generates time translations in the asymptotically flat region), and let $Y$ be the Killing vector which defines axisymmetry, with $2\pi$-periodic orbits. Given a connected component $\mcH_{\red{(i)}}$ of the event horizon, one then has the decomposition (cf., e.g., \cite{SudarskyWald93})
\begin{equation}\label{30IV23.1a}
  X =  X_{\red{(i)}} - \Omega_{\red{(i)}} Y
   \,,
\end{equation}
where $X_{\red{(i)}}$ is a globally defined Killing vector which is  tangent to the generators of $\mcH_{\red{(i)}}$. The constant $\Omega_{\red{(i)}}$ is called the \emph{angular velocity} of $\mcH_{\red{(i)}}$.
One then defines the \emph{surface gravity} of $\mcH_{\red{(i)}}$ by the formula
\begin{equation}\label{30IV23.1b}
 \frac{1}{2} \nabla \big(
  g(X_{\red{(i)}},X_{\red{(i)}})
   \big)
     =  - \kappa_{\red{(i)}} X_{\red{(i)}}
      \,.
\end{equation}
So the starting point for these definitions is the hypothesis of existence of at least two Killing vectors. But the second Killing vector, $Y$,  is not available \emph{a priori} when  attempting to classify stationary black holes.

It turns out that, under the hypothesis of $I^+$-regularity, both $\Omega_{\red{(i)}}$ and $\kappa_{\red{(i)}}$ can be defined \emph{without assuming existence of a second Killing vector}, in particular without assuming analyticity. The point is that the arguments in~\cite{HIW,ChAscona,ChNguyen,IM,ChCo} can be carried-out in a way where analyticity is
only used to translate the properties of the Taylor series of the metric at the event horizon to the whole spacetime. Here a useful observation is that, under our hypotheses, the sections of the event horizon are spheres, and therefore every Killing vector of the metric on the space of generators is periodic.
This implies that we can translate the current problem to that considered in \cite{IM}, of a compact null hypersurface with a global cross-section and with closed generators.
It follows,
by the arguments in  these references, that the metric at a stationary vacuum rotating horizon has a Taylor series identical to that of a metric with two Killing vectors.
 Thus, there exists \emph{on $\mcH_{\red{(i)}}$} a decomposition similar to \eqref{30IV23.1a} of the stationary Killing vector field $X$,
\begin{equation}\label{30IV23.1a}
  X =  X_{\red{(i)}} - \Omega_{\red{(i)}} Y_{\red{(i)}}
   \,,
\end{equation}
where now $Y_{\red{(i)}}$ is a $2\pi$-periodic vector field defined near $\mcH_{\red{(i)}}$ 
which, at $\mcH_{\red{(i)}}$,  descends to a Killing vector of the metric on the space of null generators of the Killing horizon $\mcH_{\red{(i)}}$.
If $\kappa_{\red{(i)}}$ is non-zero at one point,
it is non-zero throughout $\mcH_{\red{(i)}}$.%
\footnote{In this case $Y_{\red{(i)}}$ extends
 \cite{AIK,PetersenExtension}  to a Killing vector of the spacetime metric near $\mcH_{\red{(i)}}$,
but it remains open whether or not it extends to a Killing vector defined throughout the domain of outer communications except for near-Schwarzschild solutions~\cite{AIK2,AIK3}; see
 \cite{MinguzziGurriaran,ReirisCH,PetersenCompact,PetersenRacz} for related results.}
It then follows, e.g.\ by a contradiction argument, that if it vanishes at one point, it vanishes throughout $\mcH_{\red{(i)}}$.
 Using the divergence theorem applied to the Komar integral with the stationary Killing vector $X$, together with the decomposition \eqref{30IV23.1a} near $\mcH_{(i)}$,  leads to the same Smarr-type formula as if the spacetime were stationary and axisymmetric~\cite{SudarskyWald93}:
\begin{equation}
    m =
    \sum_{\red{(i)}}
    \big(
     2 \Omega_{\red{(i)}} J_{\red{(i)}} + \frac{\kappa_{\red{(i)}}}{4 \pi} A_{\red{(i)}}
     \big)
    \,,
 \label{20IX22.1c}
\end{equation}
where $A_{\red{(i)}}$ is the area of $\hyp\cap \mcH_{\red{(i)}}$
and $J_{\red{(i)}}$ is a Komar-type integral over the intersection of $\hyp$ with $\mcH_{\red{(i)}}$:
% \ptcr{sign? coeff?}
%
\begin{equation}
J_{\red{(i)}} =  - \frac{1}{   {8\pi} } \int_{\hyp\cap \mcH_{\red{(i)}}} \nabla ^\mu Y^\nu_{\red{(i)}} dS_{\mu\nu}
    \,.
 \label{20IX22.1d}
\end{equation}

Next, suppose that $\kappa_{\red{(i)}}$ vanishes. One can then introduce the notion of near-horizon geometry of $\mcH_{\red{(i)}}$ (cf., e.g., \cite[Section~4.3.6]{ChBlackHoles}); note that the first few terms of a Taylor series of the metric at a horizon with $\kappa_{\red{(i)}}=0$ suffice for this.
The near-horizon geometry is invariant under the flow of the Killing vector  $Y$ and is therefore axisymmetric, hence arising from the near-horizon geometry of a member of the Kerr family with zero surface gravity~\cite{Hajicek3Remarks,LP1}.

Since the surface gravity is only defined a posteriori, invoking the arguments presented above, a convenient device is to introduce the \emph{mean surface gravity}, defined as follows: Near $\mcH_{\red{(i)}}$ we can introduce Isenberg-Moncrief coordinates, in which the metric takes the form
\begin{equation}\label{22X19.1}
  g =   2 (dr - r\alpha du - r\beta_A dx^A ) du
   + \gamma_{AB}  dx^A dx^B
  \,,
\end{equation}
with $\blue{\mcH_{\red{(i)}}} = \{r=0\}$.
The mean-surface gravity $\langle \kappa_{\red{(i)}} \rangle $ of $\hyp\cap \mcH_{\red{(i)}}$ is then defined as
\begin{equation}\label{1V23.11}
  \langle \kappa_{\red{(i)}} \rangle :=  \frac 1 {A_{\red{(i)}}} \int_{\hyp\cap \mcH_{\red{(i)}}} \alpha
   \, d\mu_\gamma
  \,,
  \qquad
   A_{\red{(i)}}:= \int_{\hyp\cap \mcH_{\red{(i)}}} d\mu_\gamma
   \,,
\end{equation}
where $d\mu_\gamma$ is the measure induced by $\gamma_{AB}  dx^A dx^B$ on ${\hyp\cap \mcH_{\red{(i)}}} $.

Actually, the notation $\langle\kappa_{\red{(i)}}\rangle$ is somewhat misleading
in that the  integrand   of the defining integral \eqref{1V23.11}, and hence possibly the integral, depend  upon the parameterisation of the generators of $\mcH_{\red{(i)}}$. What is only clear is that $\langle\kappa_{\red{(i)}}\rangle$ coincides with the surface gravity $\kappa_{\red{(i)}}$ when there is a  Killing vector and when the Killing flow parameterisation of the generators has been chosen. Indeed, the coordinates on $\{r=0\}$ are defined up to a rescaling and translation of the parameter $u$ along the generators,
\begin{equation}
 u|_{\mcH_{\red{(i)}}}\mapsto \bar u = e^{f(x^A,u)}u + h(x^A)
  \,,
 \label{9V22.1}
\end{equation}
%,
where $f$ and $h$ are smooth functions.
However, the sign -- or vanishing thereof -- is parameterisation independent:

\begin{Proposition}
 \label{o3V23.1}
In vacuum stationary $I^+$-regular asymptotically flat spacetimes the sign of $\langle \kappa_{\red{(i)}} \rangle$, or its vanishing, is independent of the choice of the Isenberg-Moncrief coordinates.
\end{Proposition}

\proof
Given any Isenberg-Moncrief coordinates on $\mcH_{\red{(i)}}$, we consider that coordinate transformation
\eqref{9V22.1}
which preserves the Isenberg-Moncrief form of the metric and which brings the stationary vector field $X$ to the form
$$
 X|_{\mcH_{\red{(i)}}}
 = \partial_u + \Omega_{\red{(i)}} \partial_\varphi
  \,,
$$
%,
where $\varphi$ is the usual azimuthal coordinate on $S^2$, with $\partial_\varphi \gamma_{AB}=0$.
As explained in \cite{HIW} (see Equation~(23) there),
this requires a rescaling $u|_{\mcH_{\red{(i)}}} \mapsto e^{f(x^A)}u$ under which
$$
 \alpha|_{\mcH_{\red{(i)}}} \mapsto  \bar \alpha|_{\mcH_{\red{(i)}}} := e^f
  \big(   \alpha - \Omega_{\red{(i)}} \partial_\varphi  f
  \big)
 \,.
$$
Then $\bar \alpha|_{\mcH_{\red{(i)}}} $ equals $\kappa_{\red{(i)}}$, which is constant on each  $\mcH_{\red{(i)}}$, and integration gives
\begin{equation}\label{3V23.2}
  \int_{\mcH_{\red{(i)}}} e^{-f} \kappa_{\red{(i)}} d\mu_\gamma
   =
  \underbrace{
  \int_{\mcH_{\red{(i)}}} ( - \Omega_{\red{(i)}}\partial_\varphi f
  }_{=0} + \alpha) d\mu_\gamma
   \,.
\end{equation}
Hence neither the sign, nor the vanishing of the integral of $\alpha$  depend upon the parameterisation of the generators.
\qedskip

We say that $\mcH_{\red{(i)}}$ is mean-degenerate if $\langle \kappa_{\red{(i)}} \rangle = 0$.

From what has been said it follows:

\begin{Theorem}
\label{T12X22.1b}
The near-horizon geometry of a mean-degenerate  connected component of the horizon of a stationary,  vacuum,   four dimensional, asymptotically flat and $I^+$-regular  black hole is Kerrian.
\end{Theorem}

%Recall that, under the hypotheses above, the metric induced on sections of a rotating horizon is invariant under the flow along suitably normalised generators. Since this metric is also invariant under the flow of the original Killing vector, this induced metric is invariant under  above it suffices to have one-sided-analyticity of the metric at the horizon, or any other form of unique continuation property  that symmetric Taylor coefficients in an expansion of the metric at the horizon guarantee that the symmetry extends.

%\ptc{some further inconclusive arguments in inconclusive.tex}
%\input{inconclusive}

\section{Higher dimensions?}
 \label{s9V23.1}
%\ptcr{additions and reorganisations in the section}

We end this   note with  some comments on higher dimensions.

First, the equality of the Komar mass and the ADM mass remains true in higher dimensions~\cite{BCHKK}. To make it useful in our context one needs the positive energy theorem,  the validity of which is not clear so far in   dimensions higher than seven   unless the d.o.c.\ is spin; see~\cite{Lee:book,HuangLee2,LLU,GallowayLee} and references therein.

Next,  the proof of the staticity theorem  in~\cite{Sudarsky:wald,ChWald,ChCo} generalises without further due to higher dimensions  under conditions guaranteeing that the positive energy theorem holds.

Finally, a key further step of the four-dimensional argument is the proof of existence of at least two Killing vectors. In higher dimensions this  has been established in \cite{HIW} when at least one component of the horizon is non-degenerate and rotating, assuming again that the spacetime metric is analytic.%
\footnote{We note that the argument in \cite{HIW} invokes the von Neumann ergodic theorem, but a more direct way to proceed is via a suitable version of Birhoff's ergodic theorem~\cite{LesigneBirkhoff}.}
Some partial results in the rotating-and-degenerate case are available (see~\cite{HollandsIshibashi2,KunduriLuciettiLRR} and references therein), but the symmetry-enhancement issue is not settled so far.

%\ptc{transformation formulae out}
%\reject
%\input{NonlinearGauge}
%\input{NonlinearGaugeWrong}

\bigskip

{\noindent \sc Acknowledgements:}
Useful discussions with Maciej Kolanowski are acknowledged.

%
%\bibliographystyle{amsplain}%\bibliographystyle{/usr/local/lib/texmf/bibtex/bst/amsplain}
%%\bibliography{$HOME/prace/references/hip_bib,%
%%$HOME/prace/references/reffile,%
%%$HOME/prace/references/vienna,%
%%$HOME/prace/references/newbiblio,%
%%$HOME/prace/references/newbiblio2,%
%%$HOME/prace/references/netbiblio,%
%%$HOME/prace/references/bibl,%
%%$HOME/prace/references/howard}
%\bibliography{../references/hip_bib,%
%../references/reffile,%
%../references/newbiblio,%
%../references/newbiblio2,%
%../references/besse2,%
%../references/bibl,%
%../references/howard,%
%../references/bartnik,%
%../references/myGR,%
%../references/newbib,%
%../references/Energy,%
%../references/chrusciel,%
%../references/dp-BAMS,%
%../references/prop2,%
%../references/netbiblio,%
%%../references/newbiblio4,%
%../references/besse}
%\end{document}

\bibliography{RemarkDegenerate-minimal}
\end{document}